\documentclass{desyproc}

\begin{document}
%------------------------------------
\title{\vspace{-2.05cm}
\hfill{\small{DESY 13-160}}\\[1.27cm]
What's new in ALPS-II}

%for single authors the superscripts are optional
\author{{\slshape Babette D\"obrich$^1$, for the ALPS-II collaboration}\\[1ex]
$^1$Deutsches Elektronen-Synchrotron (DESY), Hamburg, Germany}

% if the proceedings are available online (e.g. at Indico)
% please enter the contribution ID or file_name below for the DOI
%\contribID{32}
\contribID{doebrich\_babette}

% TO THE CONFERENCE EDITORS: 
% please update the following information      
% before sending the template to the authors
% \confID{800}  % if the conference is on Indico uncomment this line
\desyproc{DESY-PROC-2013-XX}
\acronym{Patras 2013} % if you want the Acronym in the page footer uncomment this line
\doi  % if there is an online version we will register DOIs

\maketitle

\begin{abstract}
This proceedings contribution gives a brief experimental update of the 
`Any light particle search (ALPS) -II' at DESY which will be sensitive
to sub-eV, very weakly coupled particles beyond the Standard Model.
First data on hidden sector photon parameter space through photon-hidden photon
oscillations in vacuum is expected in 2014.
Axion-like particle search (implying the installation of superconducting HERA magnets) could be realized in 2017.
\end{abstract}

\section{ALPS-II setup and goals review}

ALPS-II is an experiment of the light-shining-through-a-wall (LSW) type \cite{Redondo:2010dp}, and succeeds
the experiment concluded in \cite{Ehret:2010mh}. In brief, the concept is to keep 
a large number of $\mathcal{O}$(eV) photons stored in an optical cavity before a `wall', i.e., a light-proof
environment. Measuring photons beyond that wall would indicate beyond-Standard Model (BSM)
physics\footnote{The contribution to such a process
through a Standard Model background of neutrinos is negligible due to the large mass of the 
involved mediators.}: 
If the photons are stored in a magnetic field in vacuum, the BSM process could be
due to photons oscillating into axion-like or minicharged particles: Axion-like particles
have a coupling to photons similarly to the QCD axion, but a relaxed mass-coupling relation;
minicharged particles are electrically fractionally
charged fermions or bosons, arising typically in hidden-sector models. 
Even without magnetic field, the 
photons could have oscillated into hidden photons (kinetically mixed, massive extra U(1) gauge bosons)
cf., e.g., \cite{Jaeckel:2010ni,Baker:2013zta}. 

Note that the LSW setup is generically sensitive only to particles
with masses lower than the photon energy, i.e., to sub-eV masses in our setup. 
In addition, for maximum sensitivity, the conversion should
be coherent
such that the 200m long ALPS-II will be most sensitive below the $10^{-4}$eV regime for
axion-like particles and minicharged particles
and in between $10^{-4}$eV and 1.17eV for hidden photons getting mass
from a St\"uckelberg mechanism\footnote{The situation
in the case of mass from a Higgs mechanism is slightly more involved \cite{Ahlers:2008qc}.}.
For details on the here accessible
as well as the physically most interesting parameter space of these light BSM
particles, see \cite{Bahre:2013ywa}.

ALPS-II will boost its sensitivity to light particles mainly due to the following components:
Firstly, the resonator on the `production-side' before the `wall' will be complemented 
by a frequency-locked resonator behind the barrier (in the `regeneration-region').
This increases the probability of reconversion of the light BSM particle through
`photon-self interference' \cite{Hoogeveen:1990vq}.
Also, the power buildup and the amount of in-coupled
light will be enhanced  such that one has 150kW circulating power in ALPS-II in comparison to 1kW in ALPS-I.
Secondly, the magnetic length is
enhanced by the use of 10+10 superconducting HERA dipoles instead of only 1 dipole at ALPS-I. 
Thirdly, for single-photon
detection, a Transition Edge Sensor is employed.
Note that ALPS-II is set out to be about three orders of magnitude more sensitive than ALPS-I
in the search for axion-like particles and about two orders of magnitude for hidden photons.

The experiment is structured in three phases. In ALPS-IIa (ongoing), meeting the optics and detector
experimental challenges at a 10m+10m setup (cavity length before and after the `wall',
respectively) are addressed and a search for hidden photons can be performed.
In addition, the magnet straightening-techniques (see below) are studied.
ALPS-IIb will show the viability of the setup at 100m+100m length in the HERA tunnel. 
Finally, ALPS-IIc
(still to be approved)
will include the HERA dipole magnets, in order to be sensitive to axion-like (ALP) and minicharged particles.

After a thorough review of the ALPS-II Technical Design Report (an excerpt is published in 
\cite{Bahre:2013ywa}) by appointed, external referees, the DESY management has approved the first two phases
of ALPS-II. In addition, an ALPS group has been established in the DESY high energy division.
The ALPS-II collaboration comprises DESY, the 
Albert-Einstein Institute (AEI) in Hanover and the University of Hamburg.

In the following, this proceedings contribution briefly updates the
status presented at last year's workshop \cite{Dobrich:2012du}.

\section{Experimental status of optics, magnets and detector}

\noindent
{\bf Optics:}

In brief, the optics challenge is due to the necessity of
frequency-locking and aligning the production and regeneration resonators whilst
requiring sensitivity to possible single photon-events from BSM physics:
To keep both cavities frequency-locked, laser light must also 
oscillate in the regeneration cavity (to stabilize the cavity,
Pound-Drever-Hall locking is employed). 

The production cavity is set out to host 35W of {\it infrared} (1064nm) light
at a power build-up  of 5000 ($\sim$ number of photons reflections inside the cavity).
Thus, to discriminate signal photons (e.g., due to axion-like particles), 
the regeneration cavity on the other hand is
locked with only a few mW of frequency-doubled, {\it green} light, from the
same laser source as the light oscillating in the production resonator. 
The power build-up for infrared light in the regeneration cavity, however,
amounts to 40000 (amplifying the signal photons). 

This setting requires great care to avoid that infrared
light enters the regeneration region when coupling the green light into 
the regeneration cavity. 
In addition, no infrared photons should 
be created from the green due to down-conversion processes.
Both these effects: light-tightness of the production region
and down-conversion effects are quantified stepwise with the integration of
components to the setup. So far no show-stopper has occurred.

On top of that, the simultaneous locking of both cavities must be shown,
this is done at a 1m test-setup at the AEI in Hannover and reported on in \cite{robin}.
Note that in principle, other locking schemes are conceivable \cite{Mueller:2009wt},
and studying both complementary methods is worthwhile.

In Hamburg, at the ALPS-IIa site (HERA West facility, one floor below ground level), 
infrastructural measures are mostly completed and
successful studies with a low-finesse cavity
have been performed throughout this year: As the mirrors of the ALPS-II resonators
will be located on different optical tables in the 10m+10m setup, such studies were necessary 
to assess whether a stable operation with the high-finesse dichroitic 
mirrors will be possible with available vibration dampening.
Our measurements show an integrated RMS noise for the free-running cavity on the order of a few 
$\times 10^{-9}$m down to 10Hz,
allowing to operate the envisaged cavities in principle.
Similar studies
to evaluate the conditions in the HERA tunnel (where 
ALPS-IIb and ALPS-IIc will be located) are in their initial phases.

To facilitate the alignment of the 2+2 cavity mirrors,
the plane, rectangular 
central mirrors will be fixated on a common, very smooth central `breadboard' \cite{Bahre:2013ywa}.
Considering the locking mechanism as described above, we have
devised a shutter box which must match the following criteria:
Sealing off the regeneration cavity from infrared photons
(except for calibration purposes) whilst allowing for an in-coupling
of the green light for locking purposes. The light-tightness
of the box (through milling of a labyrinth) 
must be realized without
interfering with the planarity of the breadboard surface.

The delivery of the entire set of the set of highly reflective mirrors with 250m ROC
is expected soon (choosing this ROC
ensures that the mirrors can be used also in the succeeding stages of ALPS-II).

\vspace{0.1cm}
\noindent
{\bf Magnets and vacuum:}

As reviewed in detail in \cite{Bahre:2013ywa}, 
achieving the foreseen power-buildups in the optical resonators
whilst using a setup with 20 magnets
requires reinstating the full aperture of the proton beam tube inside the 
HERA magnets. 
For the accelerator-use, the beam pipe was bent 
such that the free aperture amounts to $\sim$35mm, instead of $\sim$55mm.
Note that if the aperture would be not reinstated approximately,
the envisaged power buildup would 
only allow for an installation of 4+4 magnets
(a high power buildup necessitates to have
little clipping losses). However, with 4+4 dipole magnets
only, probing couplings beyond the range of
CAST \cite{FerrerRibas:2012ru} and in the physically most interesting parameter
region of ALP-photon couplings of g $\lesssim 10^{-10}$GeV$^{-1}$ would not be possible.

We have devised a method to reversibly straighten the beam pipe
to an effective aperture of $\sim$50mm by the insertion of `pressure
props' that stabilize the cold mass against the cryostat wall. This
was demonstrated
first conceptually in a non-functional `PR'-magnet and subsequently, the
deformation `props' were inserted at the ALPS-I magnet in
a dipole test bench. The magnet was quenched on purpose several times to 
demonstrate the stability of the setup in September 2012. The quench current was higher than
during the last runs of ALPS-I.

Note that ALPS-IIc will
only require to use {\it spare} dipole magnets. The dipole magnets in the HERA
ring itself will remain in place.
Straightening of the 20 dipoles for ALPS--IIc is not foreseen
before 2014. Thus, at the moment, the required tools for this procedure
are optimized. Fast surveying
techniques of the cold and warm dipole bore, respectively, are being studied.

As HERA is equipped with ion getter and titanium sublimation pumps,
detailed studies of the light emission in getter pumps have been 
started this year to asses the possibility fake-signals on our detector,
if getter-pumps are used in the regeneration region \cite{severin}.
In any case, the titanium sublimation pumps 
will suffice if a problem with light emission is inferred.

\vspace{0.1cm}
\noindent
{\bf Detector:}

For detecting photons at ALPS-II, fiber-guiding the
photons to a Transition-Edge Sensor detector (TES),
is foreseen.
In a nutshell, the current-change 
in a sophisticated multilayer superconductor, 
operated at the superconducting edge,
is picked up by an inductively coupled SQUID at cryogenic temperatures.
The cryostat used in ALPS-II to host the TES and the SQUID is an adiabatic demagnetization
refrigerator (ADR). 
Such a setup is perfectly suited for single-photon
detection due to its extremely low dark count rate.

Earlier this year, two TE sensors were lent to our collaboration from 
NIST in the US and AIST in Japan, respectively. In close coordination with
the PTB in Berlin, the detection system with these sensors is 
set up at DESY as reviewed in \cite{jan}.
Note, that in principle two channels are available
such that two TE sensors can be used in parallel 
(e.g., one could be used to record solely background events, first
background studies have been performed  \cite{jan}).

In a separate setup, we have studied the in-coupling
of an ALPS-II-like photon beam (i.e., diameter as
foreseen in experiment but different laser source) into a single-mode fiber.
Focusing on a fiber-coupler,
an efficiency of more than 80\% was achieved.
Note that for the overall detector sensitivity,
quantifying the coupling of the fiber to the TES inside the ADR
is also foreseen.

As a detector fall-back option and for calibration
purposes, the Princeton Instruments `PIXIS CCD' used in ALPS-I
has been characterized with respect to the 
detection of 1064nm photons \cite{eike}.

\section{To take home}

With (yet) little stringent
indications towards BSM physics in laboratory experiments, it
is worthwhile, besides the high-energy frontier, {\it also} to keep in mind the sub-eV scale as potential host of 
something undiscovered \cite{Ringwald:2012hr}, including Dark Matter \cite{Arias:2012az,lob}.
ALPS-II aims at exploring a large parameter space of new sub-eV particle physics
by combining idle accelerator infrastructure with pioneering optics and 
detector techniques.
In summary, we hope to report to the next, 10th PATRAS workshop 2014 at CERN
with first ALPS-IIa measurement data.

\vspace{0.1cm}

{\it The author would like to thank the workshop organizers for a topical and
motivating conference.
%in scenic surroundings. 
In addition, the author thanks
the Aspen Center for Physics (NSF Grant \#1066293) for a lab-timeout to pick up on different unfinished
projects and a quiet day to write up this note.}

% ****************************************************************************
% BIBLIOGRAPHY AREA
% ****************************************************************************

\begin{footnotesize}

\end{footnotesize}

% ****************************************************************************
% END OF BIBLIOGRAPHY AREA
% ****************************************************************************

\end{document}